\documentclass[twocolumn,showpacs,preprintnumbers,amsmath,amssymb]{revtex4}
\usepackage{graphicx,epsfig,amsfonts}
\usepackage{psfrag}

\begin{document}

\title{How bad is to be slow-reacting ?\\
On the effect of the delay in response to a 
changing environment on
a population's survival}

\author{Ioana Bena}
\email{ioana.bena@physics.unige.ch}
\affiliation{University of Geneva, Theoretical Physics Department, 
Quai E. Ansermet no. 24, 1211 Geneva 4,
Switzerland}
\author{Michel Droz}
\email{michel.droz@physics.unige.ch}
\affiliation{University of Geneva, Theoretical Physics Department, Quai E.
Ansermet no. 24, 1211 Geneva 4,
Switzerland}
\author{Janusz Szwabi\'{n}ski}
\email{janusz.szwabinski@physics.unige.ch}
\affiliation{University of Geneva, Theoretical Physics Department, Quai E.
Ansermet no. 24, 1211 Geneva 4,
Switzerland \\
and Institute of Theoretical Physics, University of Wroc{\l}aw,
Pl. M. Borna 9, 50-204 Wroc{\l}aw, Poland}
\author{Andrzej P\c{e}kalski}
\email{apekal@ift.uni.wroc.pl}
\affiliation{Institute of Theoretical Physics, University of Wroc{\l}aw,
Pl. M. Borna 9, 50-204 Wroc{\l}aw, Poland}

\date{\today}

\begin{abstract}
We consider a  simple-model population, whose individuals react with 
a certain delay to temporal variations of their habitat. 
We investigate the impact of such a delayed-answer on the 
survival chances of the population, both in a periodically 
changing environment, and in the case of an abrupt change of it. 
It is found that for population with low degree of 
mutation-induced variability,  being ``slow-reacting" decreases 
the extinction risk face to environmental changes. 
On the contrary, for populations with high mutation amplitude, 
the delayed reaction reduces the survival chances.

\end{abstract}


\maketitle
\section{Introduction}
\label{intro}

In the nowadays context of global warming and habitat destruction, there is an 
enhanced general interest in the impact of environmental 
changes on biological populations evolution. 
Despite this, little has been done from a theoretical point of view, 
in the frame of evolutionary dynamics modeling, towards a {\em systematic} approach
of the role of various elements involved in these complex circumstances on the population dynamics.
In a recent paper~\cite{nash1} we investigated systematically the role of the selection pressure and mutation amplitude,
as well as the impact of the quality and quantity of the habitat changes
on the behavior of a single-species population.

For simplicity and in order to extract the 
generic features, we considered the case of a 
{\em periodically changing environment}, as in 
Refs.~\cite{pease,lande,sznajd}. The case of an abrupt change 
in the environment was also addressed.
The mean-field level of description of the chosen model 
allowed us to put the finger, for the first time,  on  the  very origin 
of the emerging complex behavior of this highly-nonlinear system,
that is the delicate {\em interplay
between the different time-scale processes}. 
The role of the amplitude and period of the
environmental changes on the critical value of the selection pressure
(corresponding to a phase-transition ``extinct-alive" of the population)
was clarified. However, the intrinsic
stochasticity, the dynamically-built correlations between the individuals,
and the role of the mutation-induced variety in population's 
evolution cannot be appropriately  accounted for at a mean-field level. 

A more refined level of description, which is an
indivi\-du\-al-based one, was therefore also considered.
The main conclusions were that
the inherent fluctuations do not destroy the phase transition
``extinct-alive", and the mutation amplitude strongly influences 
the  value of the critical selection pressure, giving rise, in particular, 
to a diversity-induced resonance  phenomenon~\cite{gunton,jung}. 
The phase diagram in the 
plane of the selection and mutation parameters was discussed 
as a function of the environmental variation characteristics. 
In particular, an important aspect well-known to experimental 
biologists, see e.g.~\cite{cebrat}, 
was emerging naturally, namely that a small a\-mount of randomness, 
due to mutations, is beneficial 
for population's survival in the changing environment, while a too 
large amount definitely is detrimental to it. The differences 
between a smooth variation of the environment and an abrupt, 
catastrophic change were also clarified, pointing to the beneficial role 
of the mutation in ensuring species survival after a catastrophe.
 
In this short paper we shall address another aspect of this survival problem,
namely the role of the {\em delay} in the ``reactions" of the individuals. 
The lagged response to environmental changes is a phenomenon widespread 
in nature~\cite{thompson,both,anders,weatherhead,both2,hanski}. However, an extensive theoretical analysis 
of its impact on population dynamics is still lacking. The role and effects of time-delay in biologically systems has been 
addressed previously in the context of Lotka-Volterra type of dynamics 
of interacting species~\cite{delay}, where the ``delay" was included at 
the level of the coupling between the species.  
Here we are considering a  different problem, namely the 
delayed-response of the individuals of a single-species population to a 
changing environment. Using a simple model, we shall try to clarify 
the degree and limits 
of validity of the commonly-spred belief that ``a population of 
fast-reacting individuals has better survival chances face 
to changes in their environment".
   
%%%%%%%%%%%%%%%%%%%%%%%%%%%%%%%%%%%%%%%%%%%%%%%%%%%%%%%%%%%%%%%%%%%%%%%%%%%%%%%%%%%%%%%%%%%%%%%%%%%%%%%%%%%%%%%%
\section{Model}

We consider the same type of model as in Ref.~\cite{nash1}, namely 
a population of {\em hermaphrodite} individuals 
(i.e., which, although bisexual, need mating for reproduction), 
living on a two-dimensional square lattice of size $L \times L$. 
We assume that the individuals cannot cross the borders of the lattice. 
Moreover, the lattice has a {\em finite carrying capacity}, which comes
from the exclusion assumption that there is at most one individual in 
each lattice node.
 
The dynamics of the population takes place at discrete time-steps
and is the result of: natural selection (interaction with the environment), 
individual motion, mating and reproduction, as described below.\\ 

\noindent{\em A. Natural selection. Individual 
trait, time-dependent optimum, fitness, delayed-response,  
selection pressure, extinction probability.}\\  
Each individual $i$ is characterized by its {\em trait} 
or {\em phenotype}, which for simplicity is represented here
through a real number  $z_i \in$ [0,1]. 
The trait is fixed once and for all at the birth of the individual.

The population lives in an environment whose influence on the individuals is  
encoded in the value of the so-called {\it optimum},  $\varphi \,\in\,  [0,1]$, 
which we suppose to be homogeneous in space, but 
{\em periodically variable in time} $\varphi=\varphi(t)$.
Moreover, we  consider here the simplest possibility,
\begin{equation}
\varphi(t)=0.5+A \sin\left(2 \pi \;\frac{t-t_{{\rm init}}}{T_0}\right) 
\,\Theta(t-t_{{\rm init}})\,.
\label{varopt}
\end{equation}
Here $A$ denotes the amplitude of the environmental oscillation,
with $0 < A\leqslant 0.5$, $T_0$ is its period, and $t_{{\rm init}}$ 
is the moment of onset of the optimum perturbation; 
$\Theta$ is the Heaviside step function.

The case of an abrupt change in the environment,
for which the optimum jumps at $t=t_{\rm init}$ from 
$\varphi=0.5$ to $\varphi=0.5 +A$ was also considered.

An individual $i$  ``reacts" with a certain specific 
delay $\tau_i$ to the changes in the environment. 
This means that its instantaneous {\em fitness} 
(or ``adequacy to the environment", see below) 
at time $t$  is determined by the value of 
the optimum at a previous time $(t-\tau_i)$, 
\begin{equation}
f_i(t)=1-|z_i-\varphi(t-\tau_i)|\,.
\label{fitness}
\end{equation} 
The fitness determines the {\em instantaneous individual 
extinction probability per 
time step $p_i(t)$}, according to the  following expression:
\begin{equation}
p_i(t)\,=1-\,\exp\left[- \frac{{\cal S}}{f_i(t)}\right]\,,
\label{surv}
\end{equation}
where ${\cal S}$ is a parameter which models the {\em selection pressure} 
 of the environment  and constitutes a main  control parameter 
of the system. During its life-time, and individual oscillates
cyclically from being perfectly-adapted, when 
$z_i=\varphi (t-\tau_i)$, i.e., from a 
minimum possible extinction rate $p_i(t)=\exp(-{\cal S})$, to
a worse adaptation, which corresponds to $z_i \neq \varphi(t-\tau_i)$ 
and to a larger instantaneous extinction probability, and finally to
a total lack of adaptation, when $p_i(t)=1$. 
The pool of adapted individuals changes thus
at each time step.

The choice~(\ref{surv}) we made of the extinction probability and  
the implicit definition of the selection pressure parameter ${\cal S}$  
are frequently encountered
in the biological literature, see e.g.~\cite{burger}. Other choices 
and thus other ways of measuring the ``selection pressure"  
are of course possible. However, most of them can be 
mapped one onto the other and/or account for equivalent {\em qualitative} 
aspects of  the interaction between the individuals and their environment.

The  {\em individual delayed-response time $\tau_i$} 
is fixed once and for all at one individual's birth. 
We consider here a simple case when the $\tau_i$'s are random 
variables drawn from an uniform distribution within 
an interval $[0,\,T_d]$. The 
upper limit of this interval $T_d$ represents another 
control parameter of the model.
It is obvious that for a periodic variation of 
the environment only the values 
$T_d < 2 T_0$ are relevant. 

Note also that an equal delay-time
for all individuals amounts simply to a change in the 
time-origin. As such, a mean-field level description of the 
population dynamics (which is already known  as inappropriate 
for describing mutation, see~\cite{nash1}) 
will not be able to account for the effects 
of the individual delay-times on the global evolution of the population. We shall  
therefore focus exclusively on the individual-based numerical simulations.\\

\noindent{\em B. Individual motion.}\\
An individual  can move to its surroundings, and
the simplest possibility that we shall adopt hereafter
is a random-walk. Namely, 
in one time step the individual jumps on the lattice, 
from its initial location to a randomly chosen nearest-neighbor one  
(i.e., a site within the von Neumann neighborhood of the initial node), 
provided that the chosen site is empty,
and that it lies within the boundaries of the system. 
If none of the 
four first-neighbor nodes is empty, then the individual cannot move, 
and thus cannot mate (see below).\\

\noindent{\em C. Mating and reproduction. Heredity and mutation.}\\ 
Suppose an individual $i$ reaches a destination node.  
If there are other individuals (``neighbors") 
in the nearest-neighborhood of this
destination site, then the individual ``$i$" 
choses at random one of these neighbors, 
call it ``$j$", for mating. 
The pair of individuals $i$ and $j$ may then give birth to 
offsprings, which are placed at random
on the empty nodes of the joint nearest-neighborhoods of 
the two parents (that counts $6$ sites); 
therefore, the maximum number of 
offsprings of one pair of parents equals $6$.
If there is no room in this neighborhood for putting 
an offspring, then this one is not born.  

The trait of a progeny $k$ coming from parents $i$ and $j$ 
is determined by the parents' traits (heredity), 
but it can also present some ``variations" due to 
different random factors, such as recombination,  mutations, 
etc. We shall assume that
\begin{equation}
z_k = \frac{1}{2} (z_i + z_j) + m_k \,,
\label{zoff}
\end{equation}
where $m_k$  represents these variations.
It brings diversification into the phenotypic 
pool of the population and we call it conventionally {\em mutation}. 
For simplicity, we shall admit that $m_k$ is a random number,
uniformly distributed in the interval $[-{\cal M},\,{\cal M}]$, 
where $0<{\cal M}<1$ is called hereafter the {\em mutation amplitude} 
and is  a control parameter of the system~\footnote{In the biological
literature parameters analogous to ${\cal M}$ are often referred to as {\em
mutation rate}. However, because of the physical aspect ${\cal M}$
designates, the term {\em mutation amplitude} looks more appropriate to us.}.
Moreover, if Eq.~(\ref{zoff}) leads to 
$z_k >1$ or $z_k<0$, then one ``renormalizes" it
by resetting  $z_k$ to $1$, respectively $0$, 
which means simply that the trait of the individuals cannot 
overcome some fixed limits.
This choice~(\ref{zoff}) for the trait of
an offspring is often made in the biological literature~\cite{burger}.

The population dynamics is thus driven by two main ``forces"
that are acting, to some extent, 
in opposite directions: selection and mutation, 
characterized, respectively, 
through the values of the control
parameters ${\cal S}$ and ${\cal M}$.
Selection, combined with heredity, tries to 
bring the average trait close to the optimum, 
while mutation introduces diversity in the 
individual traits, and thus is broadening the 
distribution of the population's traits.\\
 
The Monte-Carlo simulation algorithm considers the individuals
distributed on the lattice nodes, the initial condition being 
represented by  their  positions, the prescribed values 
of the individual traits and delay-times. The initial $N(0)$
individuals are randomly-distributed with a mean 
concentration $c(0)=N(0)/L^2$,
and their individual traits are randomly assigned from an 
uniform distribution between $0$ and $1$.

The individuals are evolving, at discrete Monte-Carlo time steps (MCS,
defined hereafter), according to the stages {\em A--C} of the dynamics as
described above, namely:\\
{\em A.} At a given time $t$ 
an individual $i$ is picked at random,
and its extinction probability  $p_i(t)$, 
corresponding to one MCS, is determined according to 
Eqs.~ (\ref{fitness}) and (\ref{surv}).  
Then a random number $r$ is extracted from
an uniform distribution in the range $[0,\,1]$; if $r<p_i$, the individual
dies, otherwise it survives.\\
{\em B.} If it survives, the individual $i$ jumps at random to one of the empty
nearest-neighbor nodes on the lattice.\\
{\em C.} Then it possibly mates and produces offsprings.\\

If at the time  $t$ there are $N(t)$ individuals in the system, 
then the above steps {\em A--C} are repeated $N(t)$ times; 
this constitutes one MCS, the unit-time of the simulations.
Afterwards, the time is advanced by one step, $t \rightarrow t+1$, 
and the above algorithm is repeated. 

As a last remark on the model, it is known on general backgrounds~\cite{toral} 
that the system size is playing a certain role on the location of the phase transitin point, 
as well as on its ``sharpness". We used for all our  Monte Carlo simulations a system 
of $100 \times 100$ lattice sites, for which we had shown previously, 
see Ref.~\cite{nash1}, that the qualitative features of the phase diagram 
are practically not affected by finite-size effects.

%%%%%%%%%%%%%%%%%%%%%%%%%%%%%%%%%%%%%%%%%%%%%%%%%%%%%%%%%%%%%%%%%%%%%%%%%%%%%%%%%%%%%%%%%%%%%%%%%%%%%%%%%%%%%%%%%%%%%%%%%%
\section{Results}

For a periodic oscillation of the optimum we investigate the 
temporal evolution of a population starting from a given
initial concentration $c(0)$. Depending on the characteristic 
parameters, the population can evolve, on the average, either to an 
``alive phase", for which its concentration is actually 
oscillating periodically, with period $T_0/2$, around a nonzero mean value,
or can get extinct after a transient period of time.
In our previous paper~\cite{nash1} we investigated in detail the phase diagram
``extinct--alive" of the population in the plane of the control parameters
${\cal S}$ and ${\cal M}$, for different values of the characteristics 
$A$ and $T_0$ of the optimum oscillation. The same type of phase diagram was also
constructed for the case of an abrupt jump of the the optimum.

\begin{figure}
\begin{center}
\psfrag{M}{${\cal M}$}
\psfrag{S}{${\cal S}$}
\includegraphics[width=0.9\columnwidth]{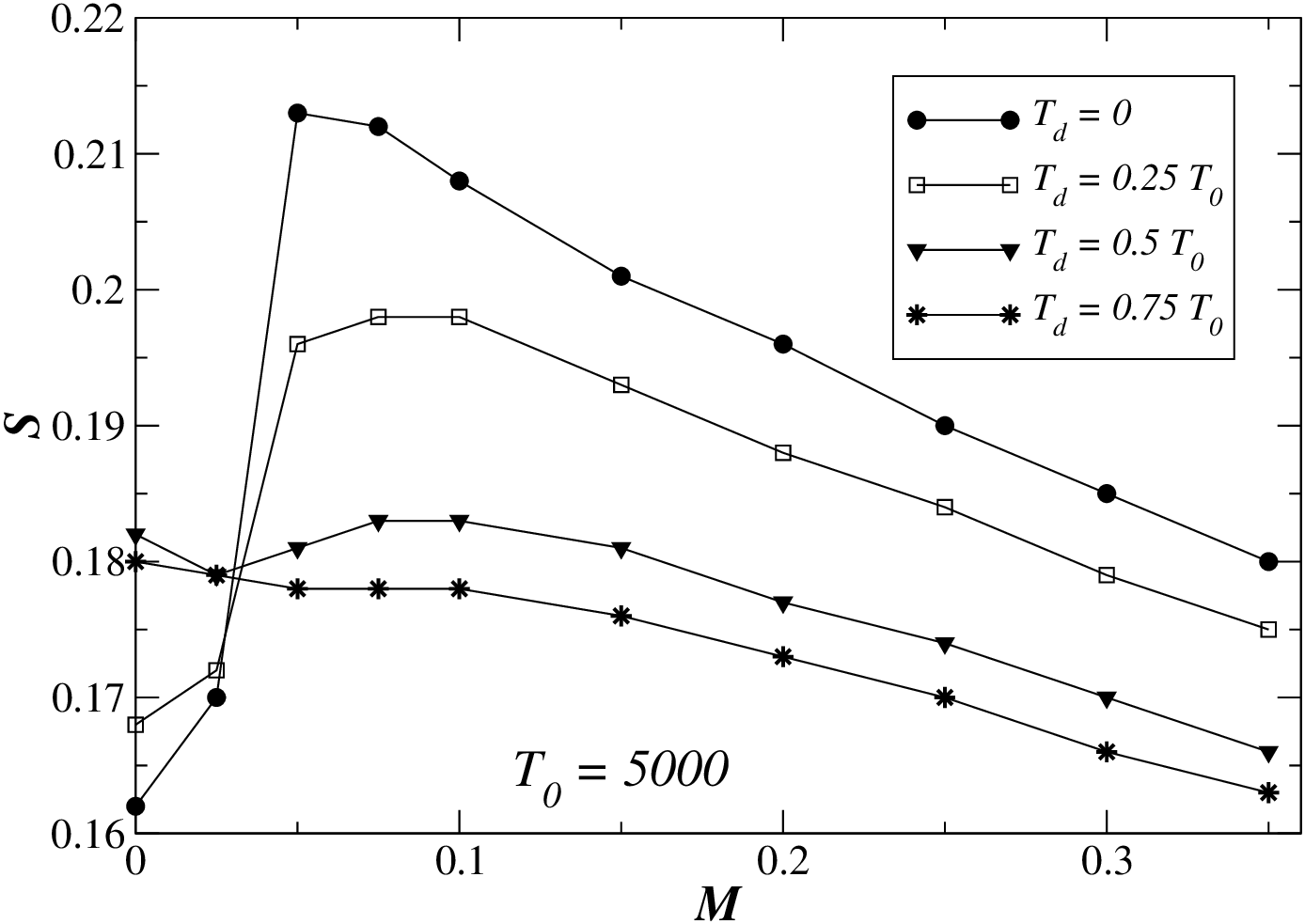}\vspace{1cm}\\
\includegraphics[width=0.9\columnwidth]{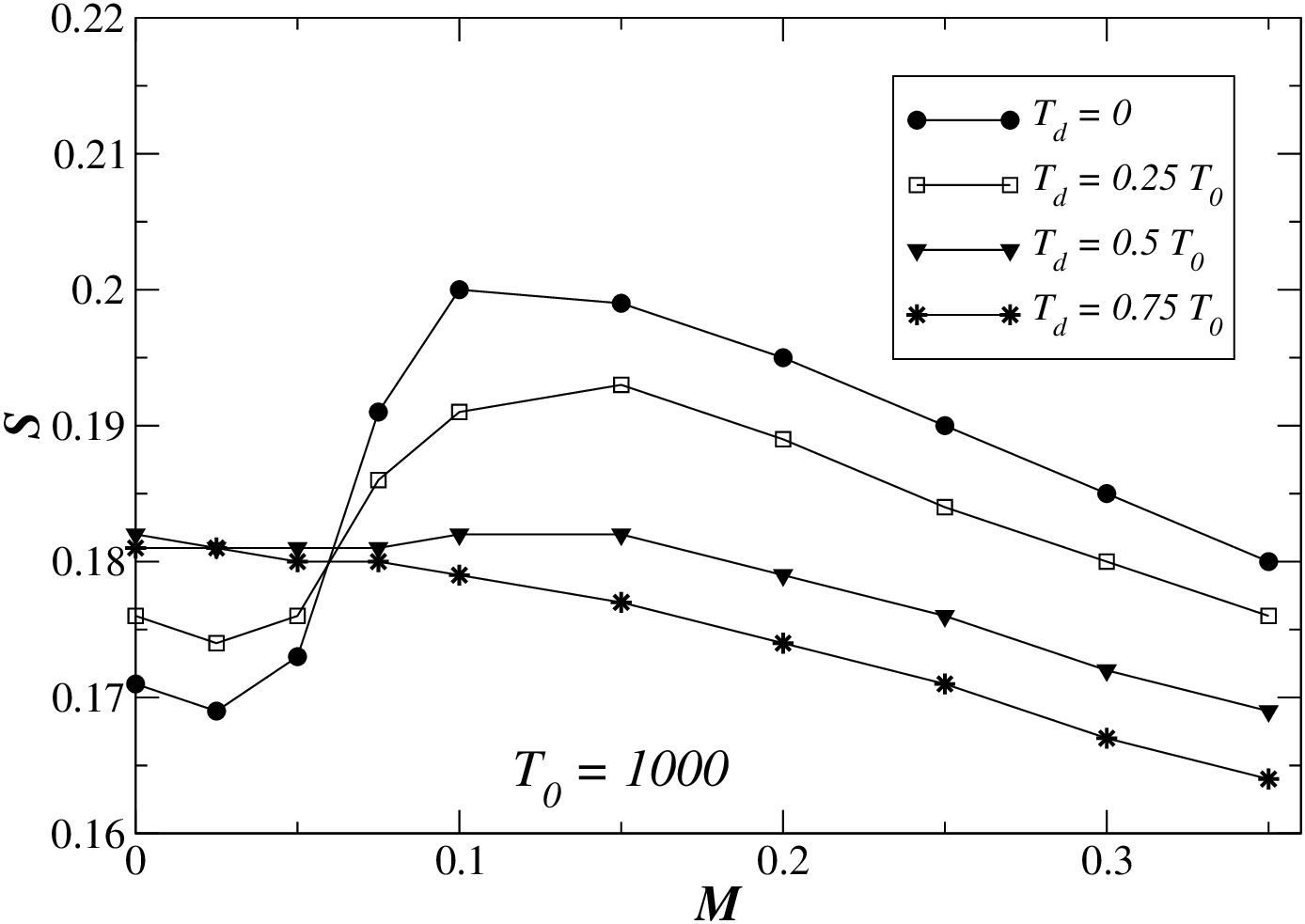}\vspace{1cm}\\
\includegraphics[width=0.9\columnwidth]{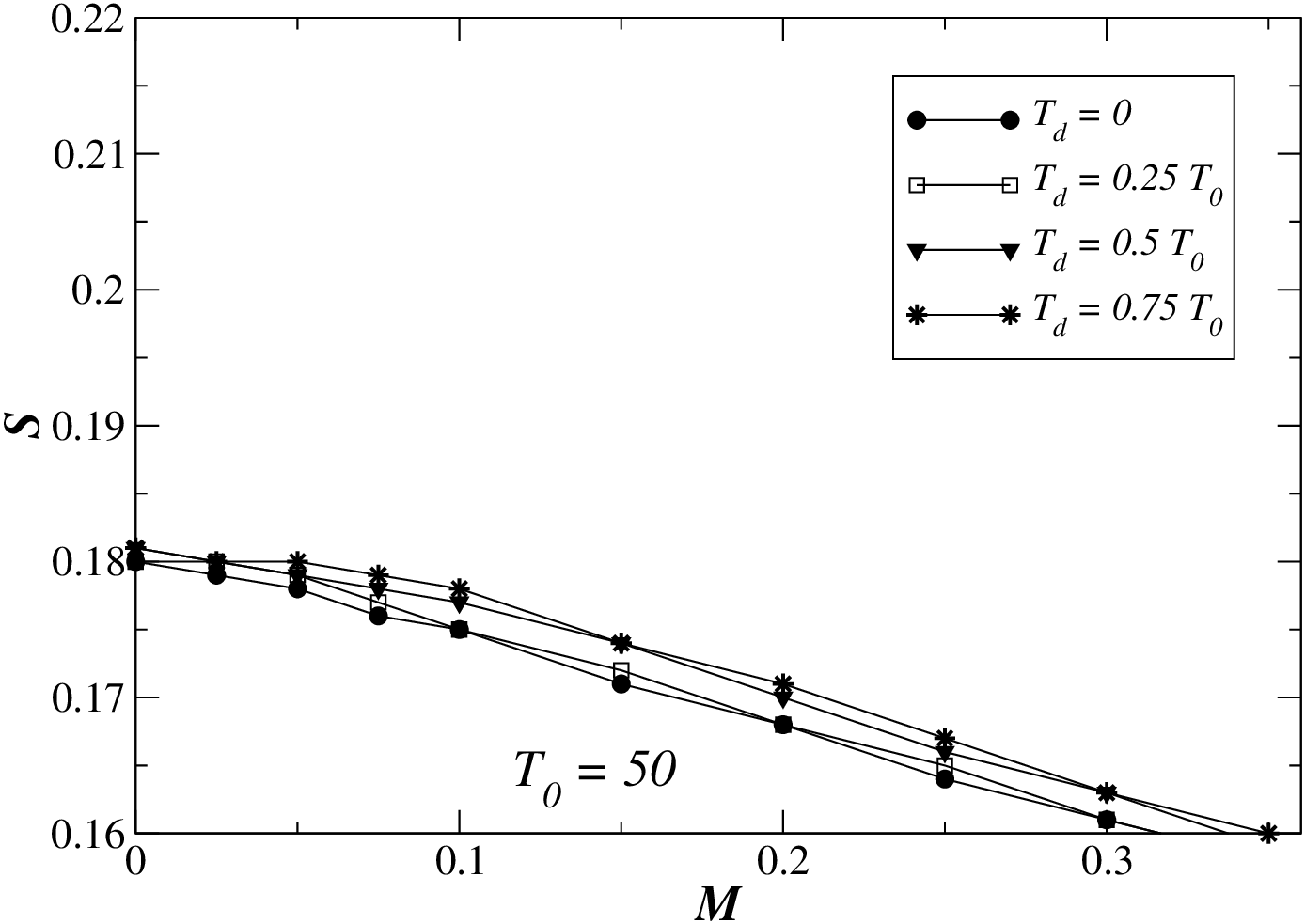}
\end{center}
\caption{Phase diagram extinct (above the curve) -- alive (below the curve) 
in the plane of the selection pressure ${\cal S}$ and mutation amplitude ${{\cal M}}$, for different values of the delay time $T_d$ and of the optimum oscillation period
$T_0$. From the upper to the lower panel, $T_0=5000, 1000$, and $50$, respectively; 
the values of the other parameters are $L=100$, $c(0)=0.7$, $A=0.3$, and 
$t_{\rm init}=1000$. 
The average was taken over 10 realisations of the stochastic dynamics and the estimated errors in the value of the critical selection pressure are less than $\pm 0.002$.}
\label{figure1}
\end{figure}

Our principal concern in this paper is to determine how the delay in the individual
response to the changing environment -- i.e., the value of the control
parameter $T_d$ -- affects the phase diagram extinct-alive of the system.
We performed extensive simulations for various range of parameters and the
main results are illustrated in Fig.~\ref{figure1}.

One notices several interesting features exhibited by these figures:\\

a) Consider first the ``intermediate" values of $T_0$ for which, as described in Ref.~\cite{nash1} for the no-delay case, one encounters the diversity-induced resonance 
phenomenon, i.e., the ``peak" in the phase-diagram illustrated in the upper and middle panels of Fig.~\ref{figure1}. Then:\\
(i) For small values of the mutation amplitude ${\cal M}$, the existence of a delay in the response of the individuals to environmental changes (i.e., $T_d \neq 0$) is {\em increasing the survival chances} of the population.  The diversity  related to the randomness in the response of the  individuals can contribute to the appearance of a larger pool of well-adapted individuals and is thus formally equivalent to an increase in the ``effective" mutation amplitude, which is beneficial for the survival~\cite{nash1}.\\
(ii) For large values of ${\cal M}$, however, adding the randomness of 
the delayed-response to the mutation-related one is leading to an 
even higher ``effective" mutation amplitude. As such, the extinction risk of the population is increased: as seen in the plots, the phase diagram for the populations with delayed-response ($T_d\neq 0$) lies always below the one of the in\-stan\-ta\-ne\-ous\-ly-reacting population ($T_d=0$).\\
(iii) Finally, the peak related to the mutation-induced diversity is generally 
still present for the systems with time delay. However, in this case the randomness in the 
delayed-response can turn a part of the pool of well-adapted individuals into 
less-adapted ones, and thus the height of the peak is reduced 
as compared to the case of an in\-stan\-ta\-ne\-ous\-ly-adapting population.
For large delays (like $T_d = 0.75 \,T_0$ in the figures) this peak can 
be even suppressed.\\

b) One concludes therefore  that the role of the delay-induced diversity is 
an increase in the ``effective" mutation amplitude. As such, it can be easily 
predicted that  for small values of $T_0$ (rapid oscillations of the environment) 
the dynamics of the system will be only slightly affected by the delay, since it is already 
only slightly sensitive to changes in ${\cal M}$. This is illustrated in the lower panel of 
Fig.~\ref{figure1}. No diversity-induced peak, i.e., no optimal ``effective" mutation amplitude is encountered in these cases, any mutation and any delay in response being harmful for the surviving of the population.\\

\begin{figure}
\begin{center}
\includegraphics[width=0.9\columnwidth]{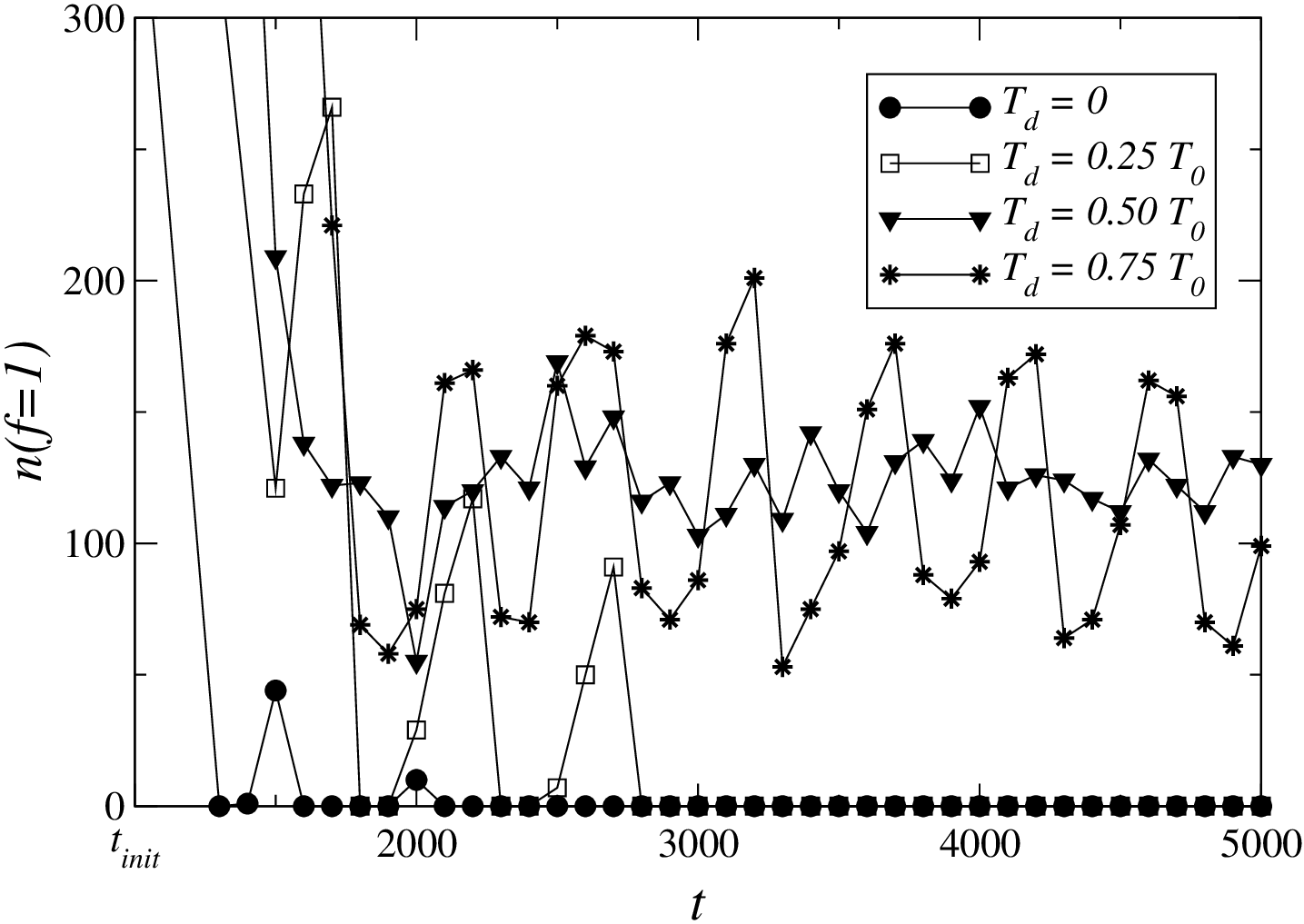}\vspace{1cm}\\
\includegraphics[width=0.9\columnwidth]{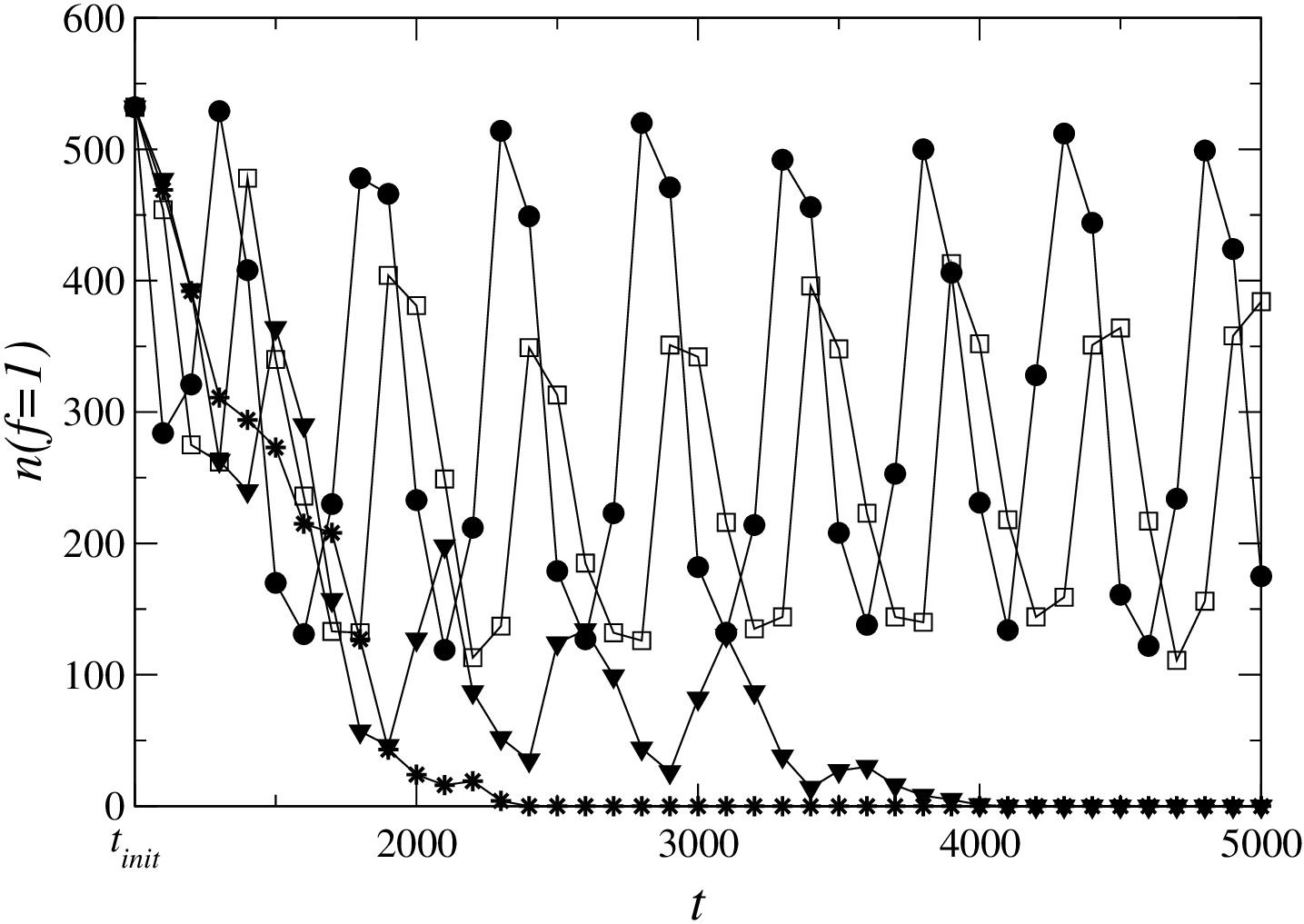}
\end{center}
\caption{Time evolution of the number $n(f=1)$ of the fittest individuals 
of a population for different values of the delay time $T_d$. Upper panel: ${\cal M}=0.025$ (small mutation amplitude),
lower panel: ${\cal M}=0.1$ (intermediate mutation amplitude). 
The legend in the upper panel also applies to the lower panel.The values of the other parameters are $L=100$,  $c(0)=0.7$, $A=0.3$, $T_0=1000$, and $t_{\rm init}=1000$, corresponding to the middle panel of Fig.~\ref{figure1}. }
\label{figure2}
\end{figure}

A way to get a better insight into the reasons for this behavior is the monitoring of the 
temporal evolution of the pool of fittest individuals (i.e., the individuals 
with $f_i(t)=1$). Figure~\ref{figure2} illustrates this point, for a fixed value of $T_0$
and three values of $T_d \neq 0$, coresponding to the middle panel of Fig.~\ref{figure1}.
The upper panel of Fig.~\ref{figure2} pertains to the region of small mutation amplitudes in the phase diagram, for which a delayed-response enhances the survival chances. The lower panel refers to the region of the peak in the phase diagram, for which delay increases the extinction risk.

One can see that for the surviving populations the number $n(f=1)$ of the instantaneously fittest individuals is oscillating periodically in time (but never reaching zero), while it 
decays (with oscillations) to zero for the populations that will get extinct. The pool of the fittest individuals is enhanced by the delay-induced diversity in systems with small mutation amplitude (upper panel of Fig.~\ref{figure2}) and, on the contrary, it is depleted by the delayed-response in populations with
intermediate and large mutation amplitude (lower panel of Fig.~\ref{figure2}).

\begin{figure}
\begin{center}
\psfrag{M}{${\cal M}$}
\psfrag{S}{${\cal S}$}
\includegraphics[width=0.9\columnwidth]{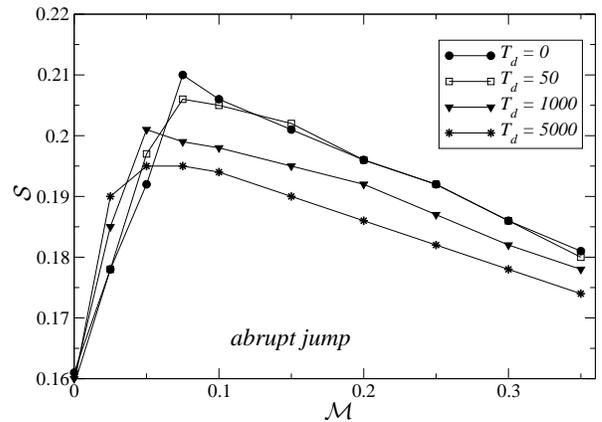}
\end{center}
\caption{The phase diagram extinct (above the curve) -- alive (below the curve) 
in the plane of the selection pressure ${\cal S}$ and mutation amplitude ${{\cal M}}$, for an abrupt jump in the value of the optimum, from $\varphi=0.5$ to $\varphi=0.8$, for different values of the delay time $T_d$. 
The values of the other parameters are $L=100$, $c(0)=0.7$, $t_{\rm init}=1000$.
The average was taken over 10 realisations, and the estimated errors
in the value of the critical selection pressure are less than $\pm 0.002$.}
\label{figure3}
\end{figure}

Finally, we addressed also the effects of a delayed 
answer in the case of a catastrophic, abrupt change 
in the environment. As illustrated in Fig.~\ref{figure3}, 
one encounters the same type of phenomena as in the case of a 
smooth variation of the optimum, namely the fact that for small 
mutation rate the largest the delay parameter $T_d$, the bigger the survival chances of the population.  

\begin{figure}
\begin{center}
\includegraphics[width=0.9\columnwidth]{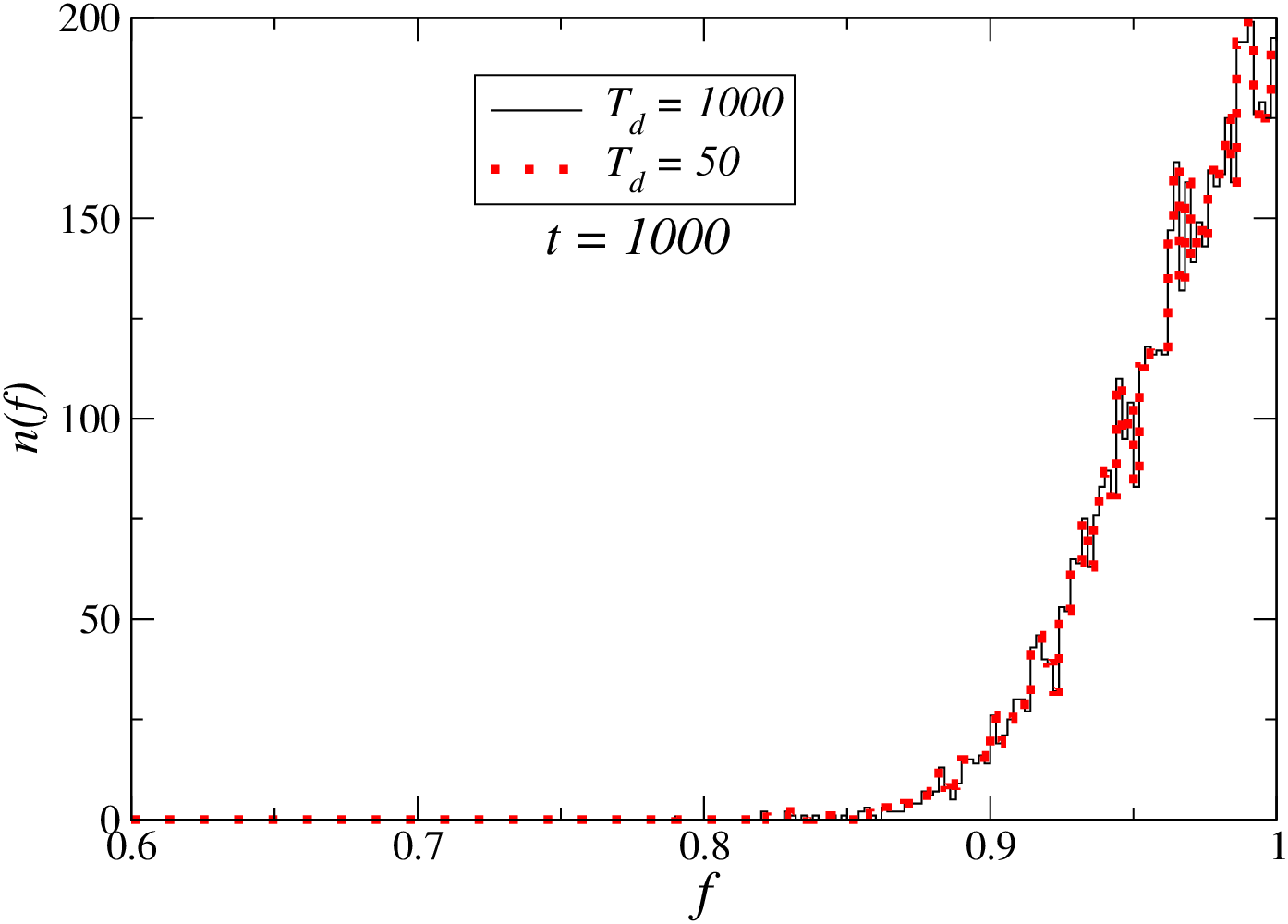}\vspace{1cm}\\
\includegraphics[width=0.9\columnwidth]{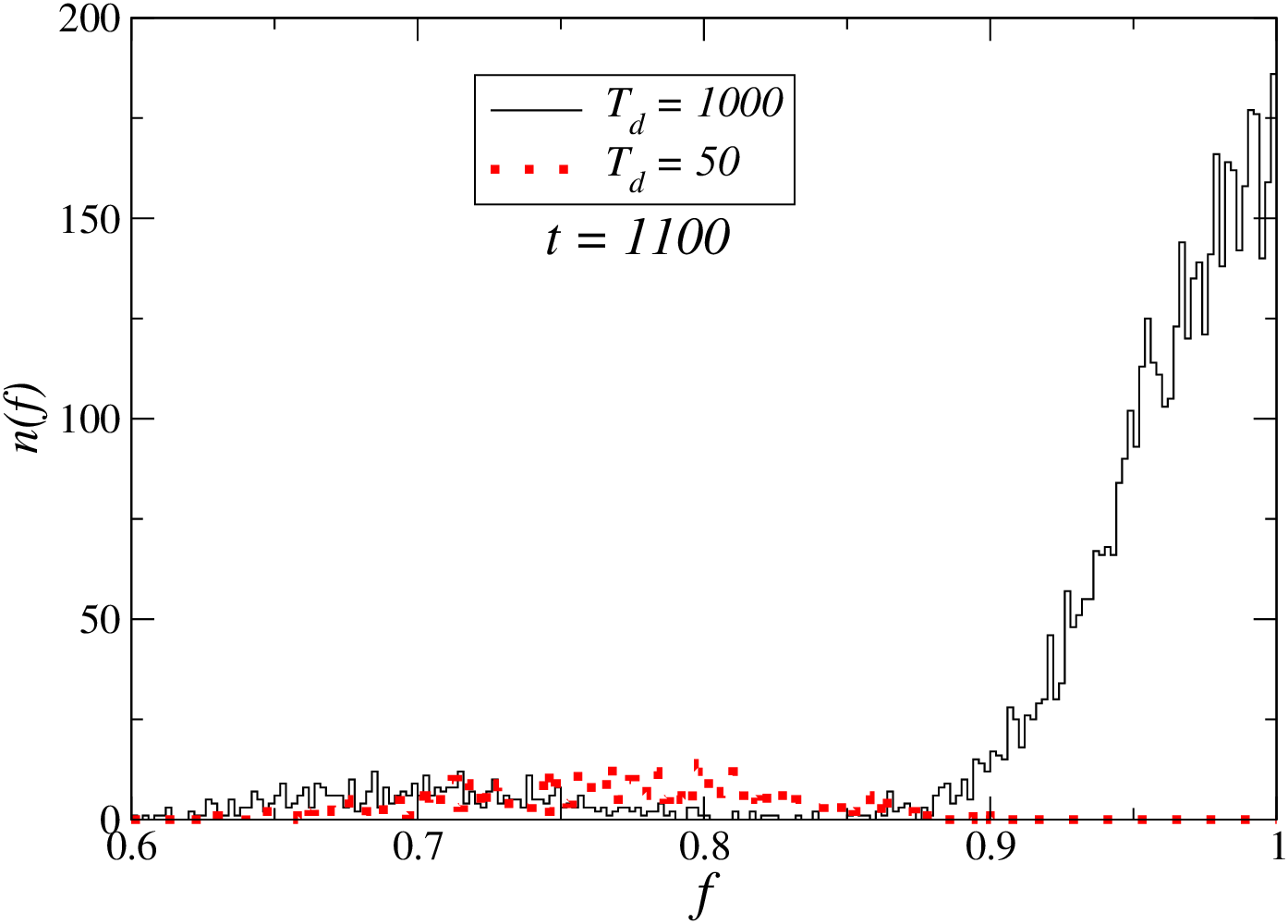}\vspace{1cm}\\
\includegraphics[width=0.9\columnwidth]{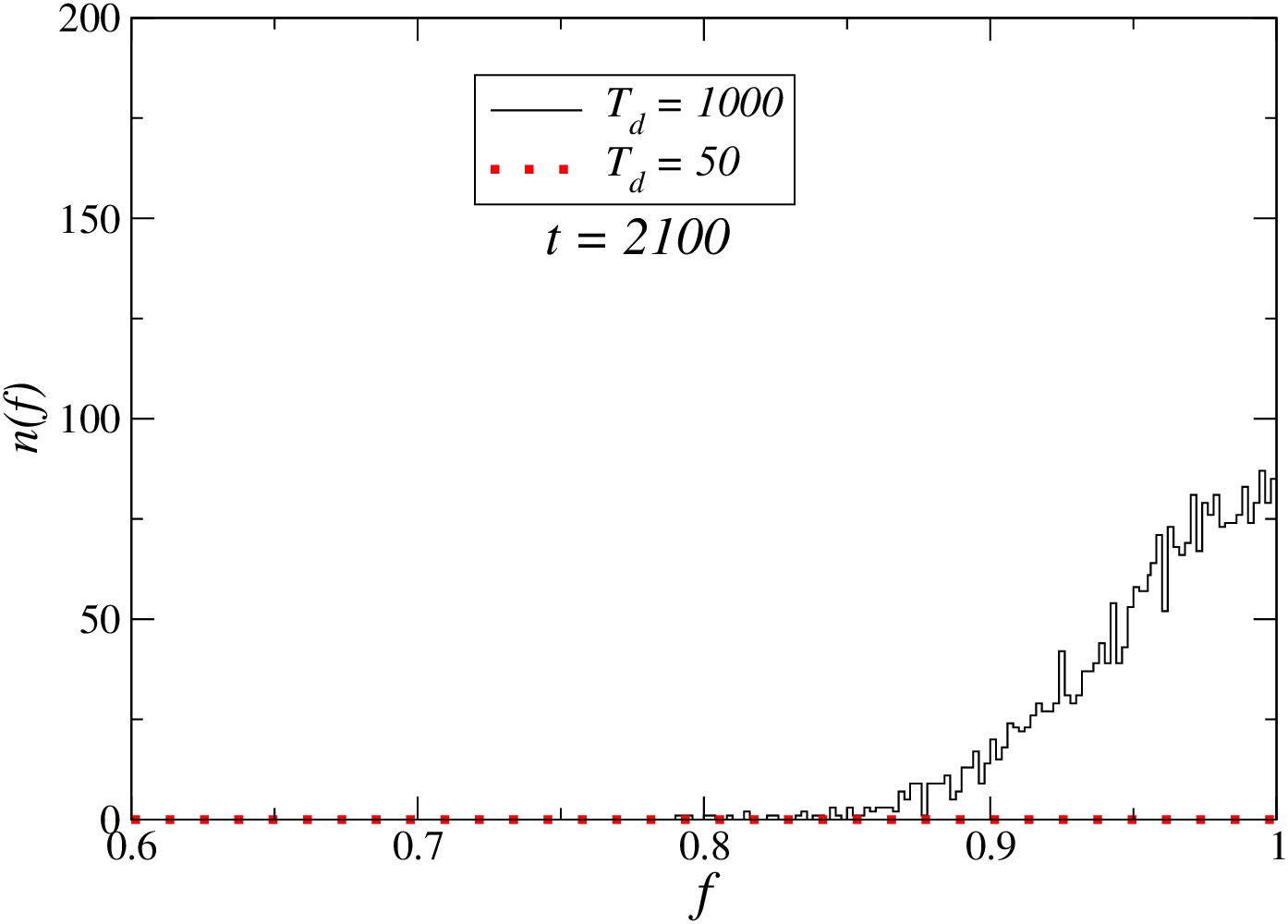}
\end{center}
\caption{Fitness histograms: number of individuals $n(f)$ versus fitness $f$
for two populations with low mutation amplitude ${\cal M}=0.05$ and different delay-response parameters $T_d=1000$ (continuous line) and $T_d=50$ (dotted line)  in case of a catastrophic event. The optimum jumps at $t_{\rm init}=1000$, from  $\varphi=0.5$ to $\varphi=0.8$. The upper panel corresponds to time $t=1000$, just before the optimum jump. The middle panel corresponds to $t=1100$, and the lower panel to $t=2100$ (when only the population with $T_d=1000$ survived). The  other parameters are $L=100$ and $c(0)=0.7$. Single runs were considered.}
\label{figure4}
\end{figure}

In order to understand the mechanism underlying  this behavior of the populations with small mutation amplitude ${\cal M}$, it is useful  to follow the temporal evolution of the fitness histogram ``number of individuals $n(f)$ versus fitness $f$". This is done in Fig.~\ref{figure4} for two populations that differ only through the value of the delay parameter $T_d$, such that one of them gets extinct, while the other one survives after the catastrophe. 
Before the catastrophe, the histogram had an important peak at $f=1$, and a tail (due to mutations) to low-fittnesses. After the catastrophe, a new peak of low-fitted individuals appears, such that the histogram becomes bimodal. One notices that the existence of a larger delay time ensures the persistence of a sufficiently large pool of high-fitted individuals even after the carastrophe, and this pool will ensure the survival of the species till the new-born individuals get adapted  slowly, through small mutations, to the new environment. 
For a surviving population the histogram becomes peaked again,in the long run,
around $f=1$. A shorter delay time $T_d$, however, cannot ensure this persistence of the high-fitted individuals pool for a  long enough  time, and the population dies, since the adaptation through mutations is not rapid enough. 

As seen in Fig.~\ref{figure3}, on the contrary, for large mutation amplitudes the larger the delay $T_d$, the higher the extinction risk, since, as in the case of a periodically-varying environment, in this case the delay-induced stochasticity adds up to the mutation,
leading to an even higher effective mutation amplitude, which is harmful for the system.

\section{Conclusions}

We  considered a  simple model of single-species population dynamics in a changing environment and we investigated the role of a delayed answer of the individuals to these habitat changes. In the case of a smooth variation of the environment, it was found that,
in general, for populations with small mutation amplitudes it is more beneficial, in terms of the survival chance, to be slow-reacting than to answer instantaneously to the variations of the environment. However, for intermediate and large mutation amplitudes, faster reactions are preferable to slower ones. In case of a very-rapidly oscillating environment,
the rapidity of reaction influences only slightly the survival chances.
The same type of statements holds true for the case of a catastrophic, abrupt jump
in the optimum.
 
As such, one has to be rather cautious with ``common-sense" statements of the kind
``a population of fast-reacting individuals has better survival chances face 
to changes in their environment". Of course, more complex and realistic models than the one we presented here are needed in order to make more detailed {\em quantitative} statements and reliable predictions for real biological systems, and to investigate further aspects of the intricate problem of a population evolving in a changing environment.\\
\\

M.D. and I.B. acknowledge partial support from 
the Swiss National Science Foundation. 
M. D. and J. S. acknowledge the COST10-SER-No.C06.0027
program for support.

\end{document}